\documentclass[conference]{IEEEtran}
\IEEEoverridecommandlockouts
\usepackage{cite}
\usepackage{amsmath,amssymb,amsfonts}
\usepackage{algorithmic}
\usepackage{graphicx}
\usepackage{textcomp}
\usepackage{xcolor}

\usepackage{hyperref}
\hypersetup{
    colorlinks=true,
    linkcolor=blue,     
    citecolor=blue,     
    filecolor=blue,     
    urlcolor=blue       
}

\def\BibTeX{{\rm B\kern-.05em{\sc i\kern-.025em b}\kern-.08em
    T\kern-.1667em\lower.7ex\hbox{E}\kern-.125emX}}
\begin{document}

\title{Interpreting Graphic Notation with MusicLDM: An AI Improvisation of Cornelius Cardew's Treatise\\

\thanks{This work was supported by IRCAM and Project REACH (ERC Grant 883313) under the EU’s Horizon 2020 programme.}
}

\author{
\IEEEauthorblockN{Tornike Karchkhadze$^{\dag}$}
\IEEEauthorblockA{\textit{University of California San Diego} \\
San Diego, USA  \\
tkarchkhadze@ucsd.edu}
\and
\IEEEauthorblockN{Keren Shao$^{\dag}$}
\IEEEauthorblockA{\textit{University of California San Diego} \\
San Diego, USA  \\
k5shao@ucsd.edu
}
\and
\IEEEauthorblockN{Shlomo Dubnov}
\IEEEauthorblockA{\textit{University of California San Diego} \\
San Diego, USA \\
sdubnov@ucsd.edu
}}


\maketitle


\begin{abstract}
This work presents a novel method for composing and improvising music inspired by Cornelius Cardew’s Treatise, using AI to bridge graphic notation and musical expression. By leveraging OpenAI’s ChatGPT to interpret the abstract visual elements of Treatise, we convert these graphical images into descriptive textual prompts. These prompts are then input into MusicLDM, a pre-trained latent diffusion model designed for music generation. We introduce a technique called "outpainting," which overlaps sections of AI-generated music to create a seamless and cohesive composition. We demostrate a new perspective on performing and interpreting graphic scores, showing how AI can transform visual stimuli into sound and expand the creative possibilities in contemporary/experimental music composition.
Musical pieces are available at \href{https://drive.google.com/drive/folders/1BIB7uqOXsDQmAn313ze5w0FN3JqUvm09}{https://bit.ly/TreatiseAI}.

\end{abstract}


\section{Introduction}

Cornelius Cardew’s Treatise is a landmark in the history of experimental music and graphic notation. Composed between 1963 and 1967, this work consists of 193 pages filled with abstract shapes, lines, and symbols that defy traditional musical interpretation. Treatise is not merely a score for performance and interpretation, but also can serve as an inspirational "springboard" for improvisation. Its abstract visual cues invite performers to explore spontaneous musical interpretations, promoting creative freedom while simultaneously challenging them to discover structure within its framework. Lacking any conventional notation, Treatise offers performers freedom, allowing each realization to be a unique artistic event.  For a small selection of interpretations we found on YouTube, please see the \href{https://youtube.com/playlist?list=PLFBnJMS2Dk5z6l5YWBRPjRmwRHMU7mEHd&si=I0CcBXW1CBXFz8H_}{playlist}, offering a taste of how different artists approach the work.
Musicians and composers have long been captivated by its open-ended nature and visual appeal, but a central question remains: how can these visual cues be interpreted innovatively and with structure, especially in today’s technologically advanced landscape?

In recent years, machine learning and generative models have revolutionized creative fields, including music composition. Generative models in the raw audio domain are now capable of producing music based on various modalities, such as text descriptions of genres, moods, or other attributes \cite{vandenoord2016wavenet, Mehri2016, donahue2018adversarial, dhariwal2020jukebox, agostinelli2023musiclm, MusicGen2023}. Diffusion models, in particular, have shown remarkable success in generating complex data like raw audio, with significant impacts on music generation~\cite{melechovsky2024mustango, schneider2023mousai, kong2021diffwave}. MusicLDM~\cite{chen2023musicldm}, a latent diffusion model for music generation, exemplifies this progress, allowing the creation of new musical works guided by text prompts. Other approaches focus on time-varying controls, such as synthesizing music audio from symbolic representations like MIDI \cite{hawthorne2018enabling, hawthorne2022multi}. Although effective, this method is constrained by its rigidity, offering very little room for interpretation. More sophisticated control mechanisms have been explored in works like DiffuseRoll \cite{wang2023diffuseroll}, which connects images to piano rolls, and in \cite{Benetatos-2022}, which developed an interactive system where user-drawn curves fill missing measures in monophonic pieces. Recent research has also integrated general controls such as envelope, pitch contour, and form into music models \cite{wu2023music, novack2024ditto, Min23Polyffusion}.

\begin{figure*}[t]
  \centering
  \includegraphics[width=.90\linewidth]{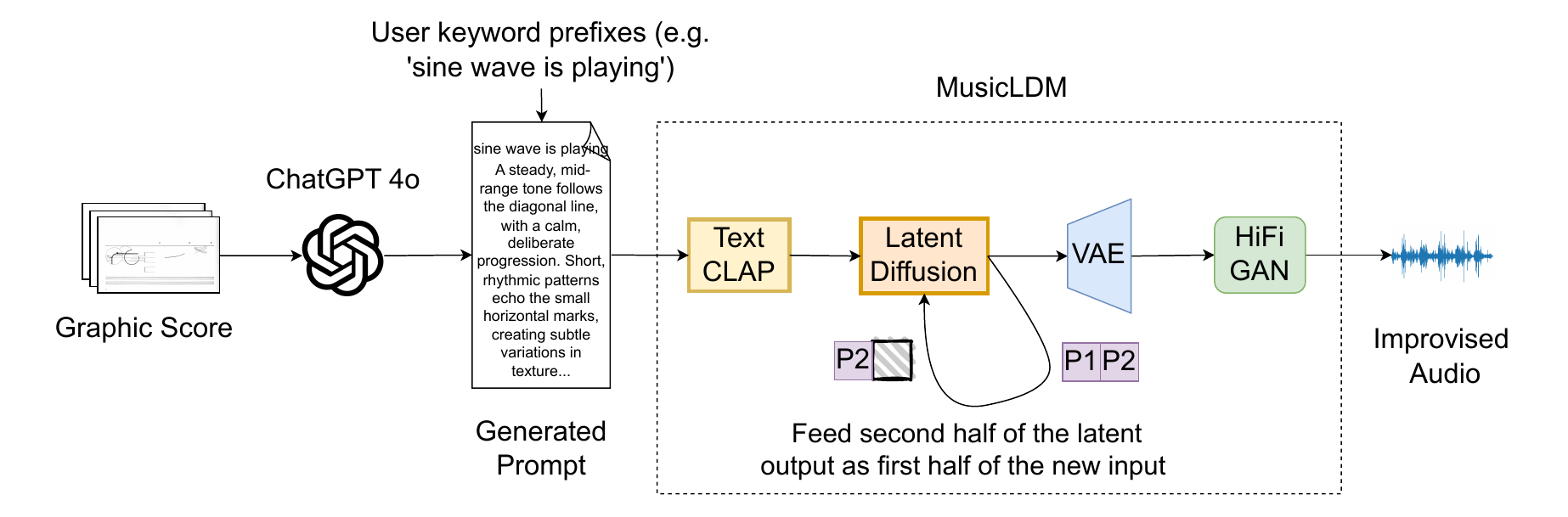}
  \vspace{-0.2cm}
  \caption{\textbf{Treatise Improvisation Pipeline:} The graphic scores are first processed by ChatGPT-4 to generate text prompts, which are then fed into MusicLDM, a latent diffusion model. Smooth transitions in the stitched audio output are achieved using the "outpainting" technique, illustrated here by the feedback loop around the latent diffusion model. Detailed explanations of this method can be found in the Methods section.}
  \vspace{-0.2cm}
  \label{fig1}
\end{figure*}

Despite these advancements, the interpreting graphic scores into coherent music remains largely unexplored in machine learning. Interpreting or improvising on a score like Treatise presents a complex challenge, as even human performers approach it with vast interpretive differences. This open-ended problem invites further exploration, offering a rich opportunity for artistic research into how abstract visual elements can be translated into musical performances through AI.

We propose a novel method for interpreting Cornelius Cardew’s Treatise as both a visual and conceptual foundation. Our approach begins by translating the visual elements of Treatise through OpenAI's ChatGPT 4o, which generates descriptive textual prompts based on selected pages of the score. These text prompts capture the nuances of the graphic elements and serve as input for MusicLDM, which generates corresponding musical sequences. 
To facilitate smooth transitions between sequences, we implement an overlapping technique with "outpainting."
In deep learning literature, "outpainting" refers to the process of extending the length of real or previously generated content. It is commonly used for tasks such as image and audio completion, as well as generating long-duration music content using diffusion models ~\cite{bar2023multidiffusion,Levy2023ControllableMP}.
Through this process, we transform Treatise's abstract visual stimuli into rich soundscapes, presenting a new dimension of interpretation.

This work showcases the application of generative AI with the graphic notation pushing the boundaries of how non-traditional scores can be engaged. By positioning AI as both a tool and collaborator, we explore the multi-modal relationship between visual art and music, offering new insights into the performance, improvisation on and interpretation of graphic scores. Please refer to Appendix \ref{setup} for the compositional setup of the AI systems and to Appendix \ref{Show and Tell} for a detailed presentation of the pieces.

\section{Method}
\label{Method}

In this section, we present our model components in the order outlined in Fig. \ref{fig1}.

\subsection{Vision Model - ChatGPT 4o}

ChatGPT, developed by OpenAI, is a cutting-edge large language model (LLM) designed to generate human-like text responses based on user prompts. Built on the transformer architecture \cite{vaswani2017attention}, ChatGPT-4 excels in a range of natural language processing (NLP) tasks, including language generation, translation, summarization, and question-answering. 
In addition to its language capabilities, ChatGPT-4 has been enhanced with multimodal abilities, enabling it to interpret both textual and visual inputs. In our workflow, we use ChatGPT to generate text that interprets Treatise scores provided as input. We then concatenate this generated text with prefix keywords that specify the overall style for the improvised audio. These keywords are incorporated to help the downstream CLAP model generate vectors that more closely align with the intended style.


\subsection{Generative model - MusicLDM and Outpainting}





Building upon the AudioLDM framework \cite{liu2023audioldm}, MusicLDM leverages denoising diffusion probabilistic models (DDPMs)~\cite{ho2020denoising, Sohl-DicksteinW15} for music generation. At the core of MusicLDM are several key components, including the CLAP encoder \cite{wu2023largescale}, a pretrained model that encodes the generated prompts from the previous section into vectors within a shared text-audio latent space. These latent vectors are then transformed into improvised audio by a downstream Latent Diffusion model, along with other vocoding components.

Although this pipeline appears comprehensive, one significant challenge arises during improvisation: stitching together the generated output. Ensuring smooth dynamics and eliminating "jumping" artifacts during post-processing is difficult. To address this, we incorporate the stitching process directly into the generation phase. In the first generation with prompt sentence, the CLAP vector $v_0 \in \mathbb{R}^{512}$ and standard Gaussian noise $\epsilon \in \mathbb{R}^{C \times T \times F}$ serve as inputs to our Latent Diffusion model $f$, producing a latent vector $z_0 \in \mathbb{R}^{C \times T \times F}$, where $C$, $T$, and $F$, correspond to channel, time, and frequency dimensions, respectively. For subsequent sentences, instead of Gaussian noise $\epsilon$, we generate smoother transitions by modifying the noise input:
\begin{align}
\epsilon' &= concat(z_k[:,T//2:,:], \epsilon[:,:T//2,:]) \\
z_{k+1} &= f(v_{k+1}, \epsilon')
\end{align}
for all $k > 0$, where k corresponds to the sequence of prompts increasing according to Treatise pages. 
This forces the Latent Diffusion model to create a smooth continuation from the previous output. We refer to this technique as "outpainting" throughout this paper.


\section{Discussion}

This work showcases the potential of AI in interpreting graphic notation, focusing on Cornelius Cardew’s Treatise. By converting abstract visual elements into music through textual prompts and using MusicLDM for sound generation, we present a novel method for translating non-traditional scores into music. Incorporating the outpainting technique allowed us to create cohesive musical compositions, offering a new perspective on interpreting such visual stimuli.


While our current approach demonstrates AI's ability to follow visual cues, it relies on text generation as an intermediate step, which may limit efficiency and control. Moving forward, we plan to replace the ChatGPT component with pre-trained visual models like CLIP. By mapping CLIP’s latent space to CLAP, we aim to create a more controllable system that bridges visual and auditory modalities, unlocking new possibilities for AI-assisted music composition and interpretation of complex visual scores.

\section{Acknowledgments}


We thank the Institute for Research and Coordination in Acoustics and Music (IRCAM) and Project REACH: Raising Co-creativity in Cyber-Human Musicianship for their support. This project received support and resources in the form of computational power from the European Research Council (ERC REACH) under the European Union’s Horizon 2020 research and innovation programme (Grant Agreement 883313). 

We also appreciate the inspiration, improvisation materials, and helpful discussions provided by Wilfrido Terrazas, Associate Professor of Music at the University of California, San Diego.

\bibliographystyle{IEEEtran}
\bibliography{treatise}

\appendices
\onecolumn

\section{Compositional Setup}
\label{setup}

\subsection{Score Interpretation with ChatGPT}

In this work, we leveraged ChatGPT 4o’s image interpretation capabilities to process Cornelius Cardew's Treatise score. We sequentially presented ChatGPT with images from pages 1 to 33 of the score, instructing it to generate four-sentence textual prompts designed to guide a music generation model. Notably, we excluded the lower horizontal lines resembling musical notation present on all pages. The generated prompts aimed to translate the abstract visual elements of the score into musical descriptions. We generated 4 prompts per page by asking ChatGPT 4o to read and describe score page form left to right. To enhance the coherence of the resulting composition, we prefixed each prompt with style-defining keywords, such as "electronic," "strings," "experimental," or "sinewave," to convey specific musical styles and moods. The selection of pages 1 to 33 was made to ensure a balanced piece length, with the cadence-like structure of page 33 providing a natural conclusion. The pages 1-7 of score used are provided in the Appendix \ref{score} with corresponding prompts in Appendix \ref{prompts}.

\subsection{MusicLDM with Overlapping Window Technique}

Our MusicLDM parameter configuration mirrors that of MusicLDM~\cite{chen2023musicldm}, operating on 10-second audio segments with a 16 kHz sampling rate, transforming them into Mel-spectrograms with dimensions of \( T \times F = 1024 \times 64 \) for time and frequency. The VAE component of MusicLDM applies a compression factor of \( r=4 \), transforming the Mel-spectrograms into latent representations where the LDM generator operates. The latent representation corresponding to 10 seconds of audio is \( z \in \mathbb{R}^{C \times T \times F} \), where \( C = 8 \), \( T = 256 \), and \( F = 16 \) are the channel, time, and frequency dimensions, respectively.
The MusicLDM and its components—including the CLAP encoder, VAE, and the HiFi-GAN vocoder—were taken from the pre-trained, publicly released checkpoint of MusicLDM\footnote{\href{https://github.com/RetroCirce/MusicLDM}{https://github.com/RetroCirce/MusicLDM}}. This checkpoint was trained on an extensive collection of music audio data.

We implement an overlapping window technique to generate continuous musical compositions. The process involves generating audio in segments, where each new segment partially overlaps with the previous one by half. The generation begins with an initial segment, followed by subsequent segments that overlap the previous ones. To ensure seamless transitions, we apply a masking strategy to the latent representation. 
The mask \( m \in \mathbb{R}^{8 \times 128 \times 16} \) retains the second half of the previous latent space as a beginning of the following vector to be generated, allowing the model to focus on generating the new, non-overlapping portion. This technique ensures a smooth musical flow, eliminating abrupt transitions or discontinuities. 

The final waveform is reconstructed by concatinating the denoised latent representations and passing them through the VAE decoder and HiFi-GAN vocoder to produce the final audio output.


\newpage
\section{Show and Tell: Analysis of Composed Tracks}
\label{Show and Tell}

In this section, we present a selection of the tracks generated using the above described method using various AI systems. Each track is accompanied by a brief descriptive explanation, highlighting the unique renderings and the key elements that differentiate them from one another. We also provide suggestions for what listeners might focus on when comparing different versions.

\paragraph{Track 1: \textit{Sinewave}}

This track was generated by providing the system with a user-defined prefix: "sine wave is playing." As we can observe, the introduction of this prefix profoundly influenced the character of the piece, making it sound akin to the tradition of experimental synthesizer music, which aligns with expectations for a Treatise-inspired work. However, the piece still retains a musical collage/montage quality that is characteristic of AI systems. The track demonstrates the system’s ability to follow the score in a meaningful way. In the first 10 seconds, the initial sparse, steady melody is quickly transformed into a more complex harmonic structure corresponding to the first page. On a larger scale, we can clearly hear how the complexity rises, then falls, stays steady, and rises again towards the end, culminating in a logical and meaningful conclusion that follows the score.

Key aspects to focus on are:
\begin{itemize}
    \item The subtle introduction of the sine wave.
    \item A notable rhythmic and complexity variation between the first and second parts (after minute 6).
    \item The return to the beginning-like steady sounds at the end, creating a meaningful conclusion.
\end{itemize}

\paragraph{Track 2: \textit{String orchestra}}

This track was generated by providing the system with a user-defined prefix: "string orchestra is playing." As we can see, the introduction of this prefix gave the piece a more classical string orchestra sound, evoking a Classical-era or somewhat cinematic character. The piece has a sharply defined collage/montage quality, often jumping from theme to theme, which is characteristic of AI systems that often operate in this manner. The track demonstrates less ability to closely follow the score and instead produces an overall sound that is more generic, though it occasionally presents interesting melodic and harmonic structures.

Key aspects to focus on are:
\begin{itemize}
    \item The repetition and transformation of many themes and melodies throughout the piece.
    \item Strong stylistic consistency with orchestral music traditions.
    \item A somewhat meaningful overall structure, with the system following the general outline of the score, and an interesting ending.
\end{itemize}

\paragraph{Track 3: \textit{Experimental}}

This track was generated by providing the system with a user-defined prefix: "Experimental music is playing." The prefix gave the piece a mixed rhythmic and experimental feel, with some noise passages interspersed between repeating rhythmic and harmonic sections. The piece also exhibits a collage/montage-like quality, though with less variability. New sounds and structures are introduced, particularly in the second part, corresponding to the score. There is a steady flow between pages 7 and 15, with increased variability from page 15 onwards.

Key aspects to focus on are:
\begin{itemize} 
    \item The transition between different types of rhythmic structures. 
    \item The overlay of form and complexity as it follows the score. 
    \item The logical fade-out ending, as suggested by the score. 
\end{itemize}

\textbf{Summary of Highlights:} Each of the selected tracks illustrates different variations of the AI system’s interpretation, from clearly defined sound texture generation to more abstractly defined styles. These tracks offer a comprehensive showcase of how the system approaches musical composition and variation, demonstrating that the system generates coherent pieces while following the score.

\clearpage
\section{Cornelius Cardew's Treatise score (pages 1-7)}
\label{score}

\begin{figure}[h] 
    \centering
    \includegraphics[width=\textwidth, height=0.42\textheight, keepaspectratio]{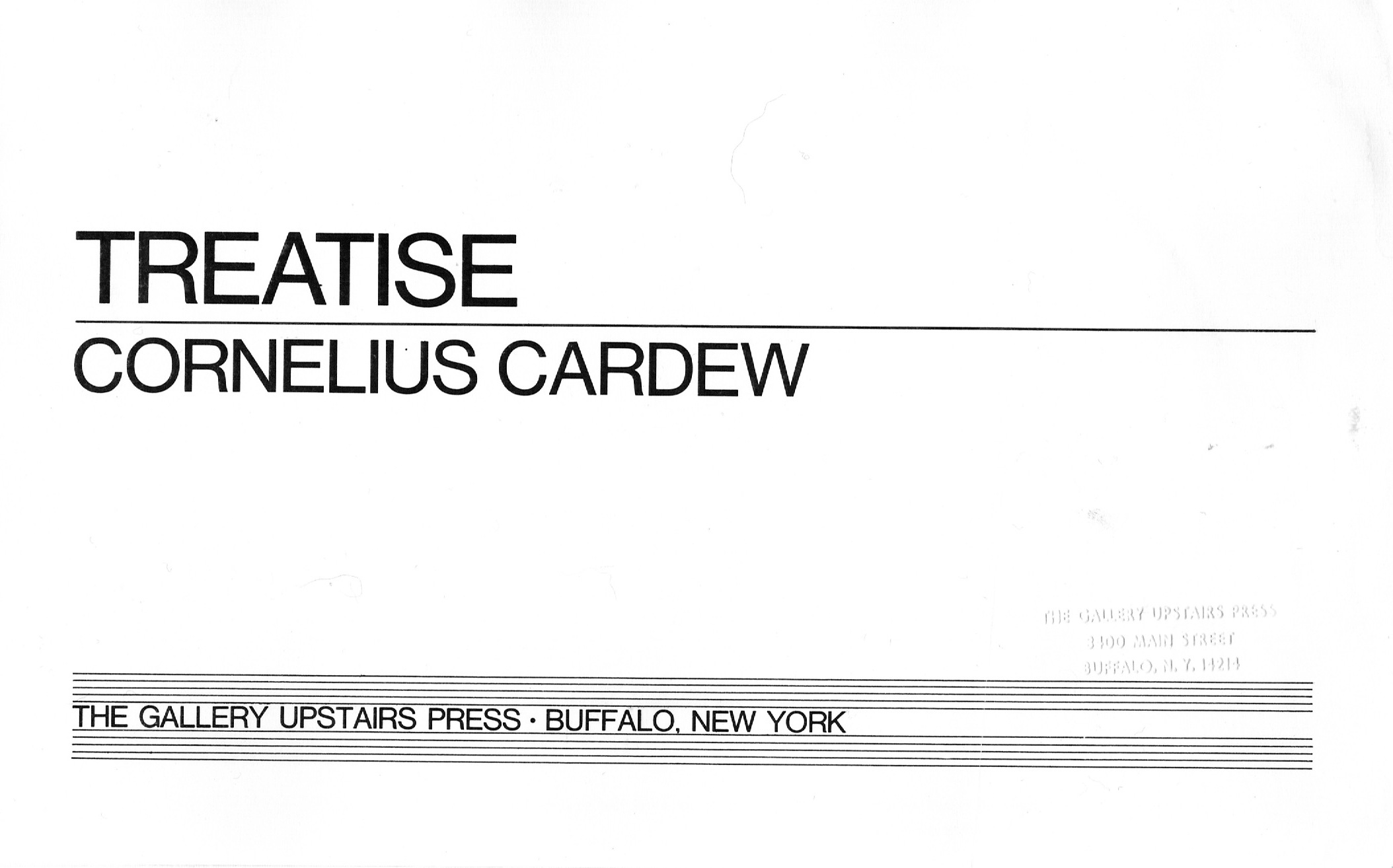}
    \includegraphics[width=\textwidth, height=0.42\textheight, keepaspectratio]{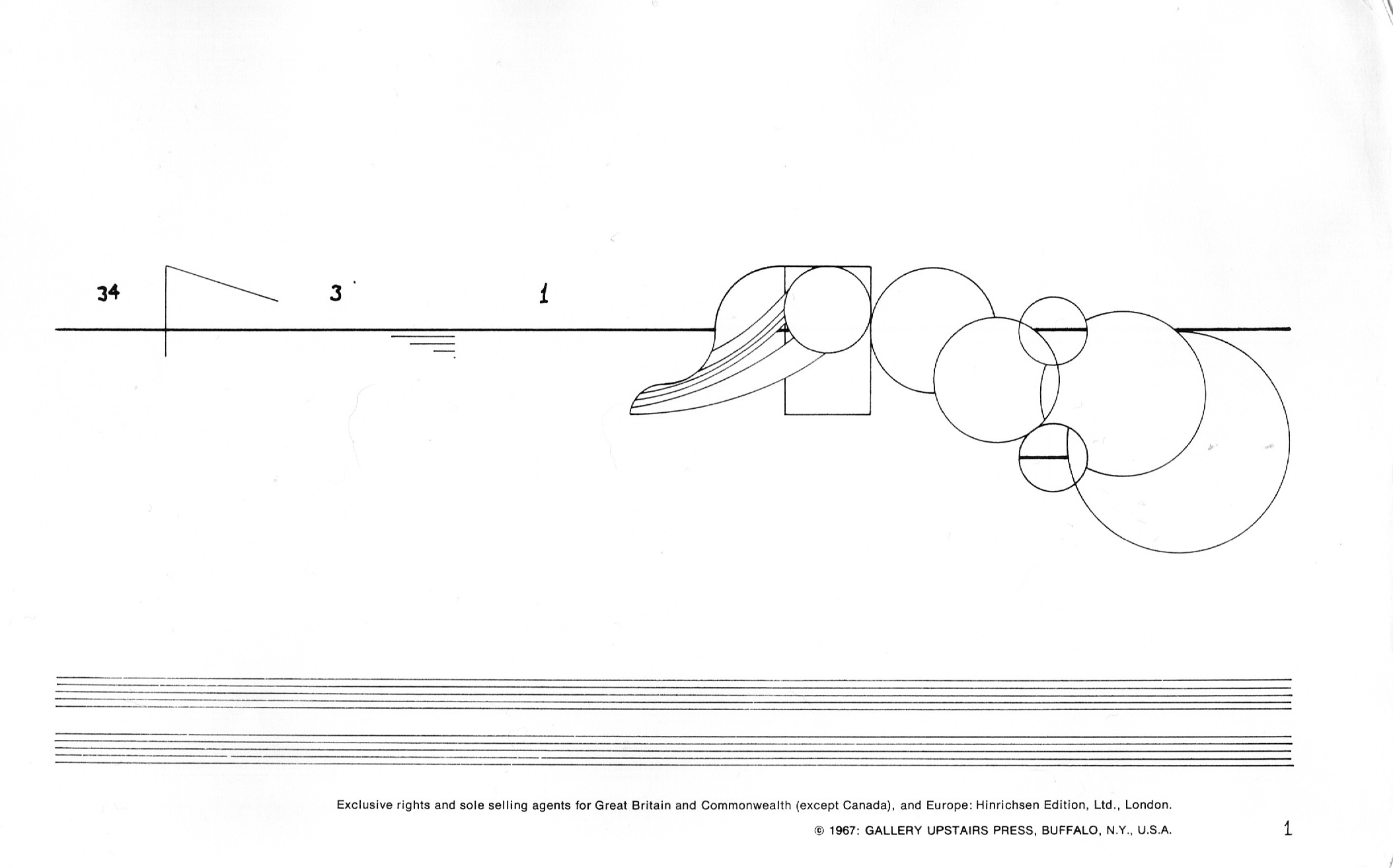}
\end{figure}

\clearpage 
\begin{figure}[!ht] 
    \centering
    \includegraphics[width=\textwidth, height=\textheight, keepaspectratio]{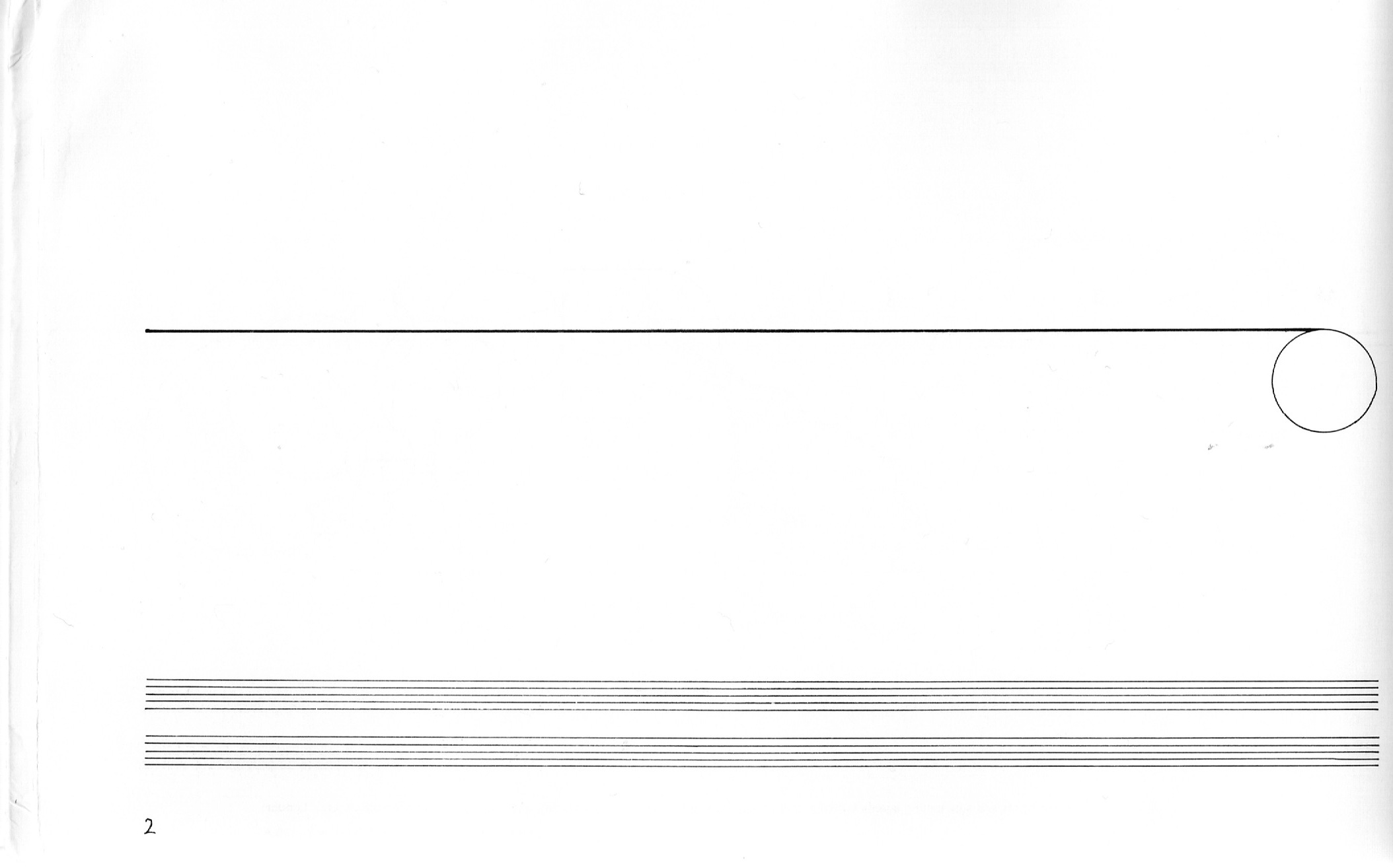}
    \includegraphics[width=\textwidth, height=\textheight, keepaspectratio]{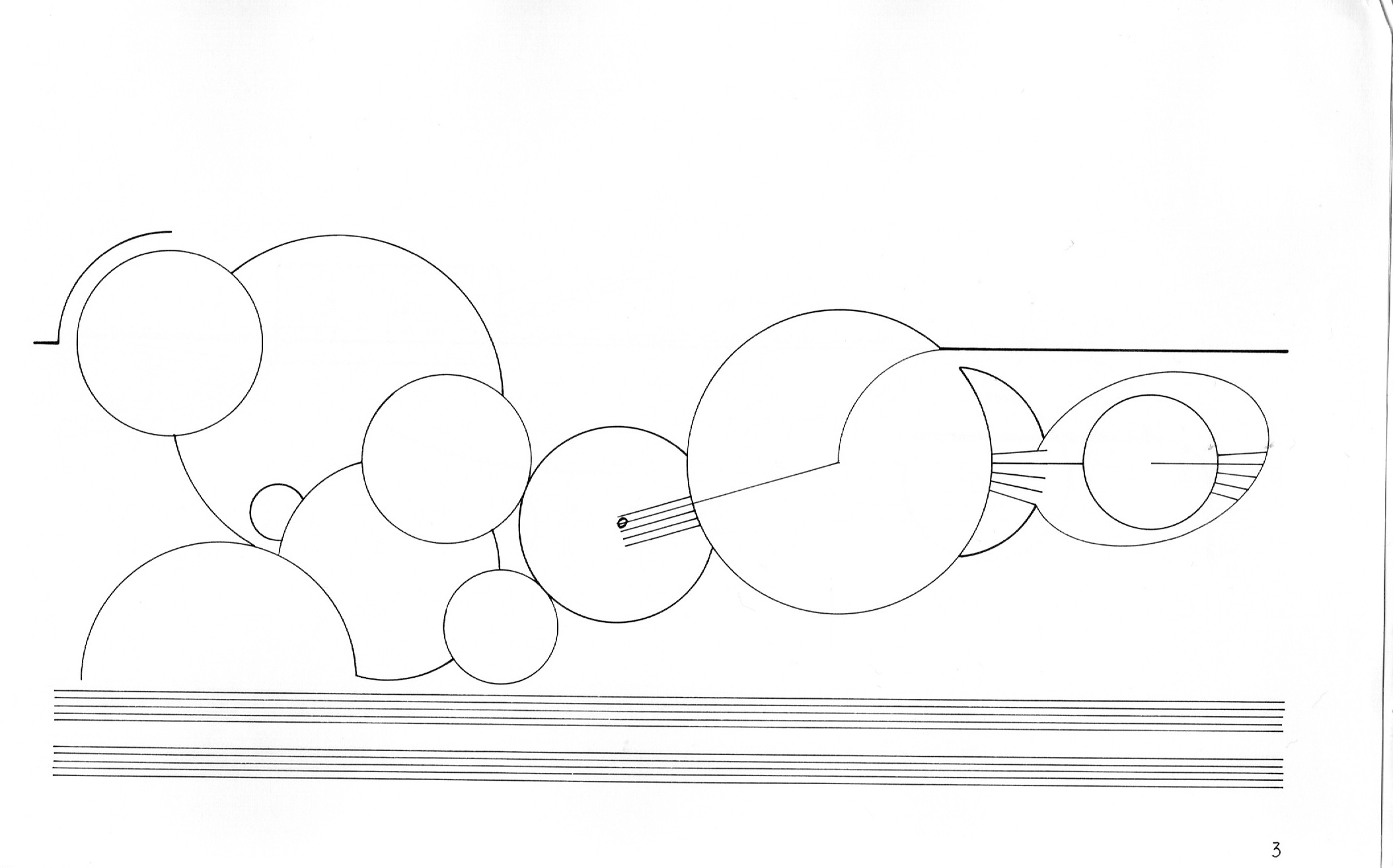}
\end{figure}

\clearpage 
\begin{figure}[!ht] 
    \centering
    \includegraphics[width=\textwidth, height=\textheight, keepaspectratio]{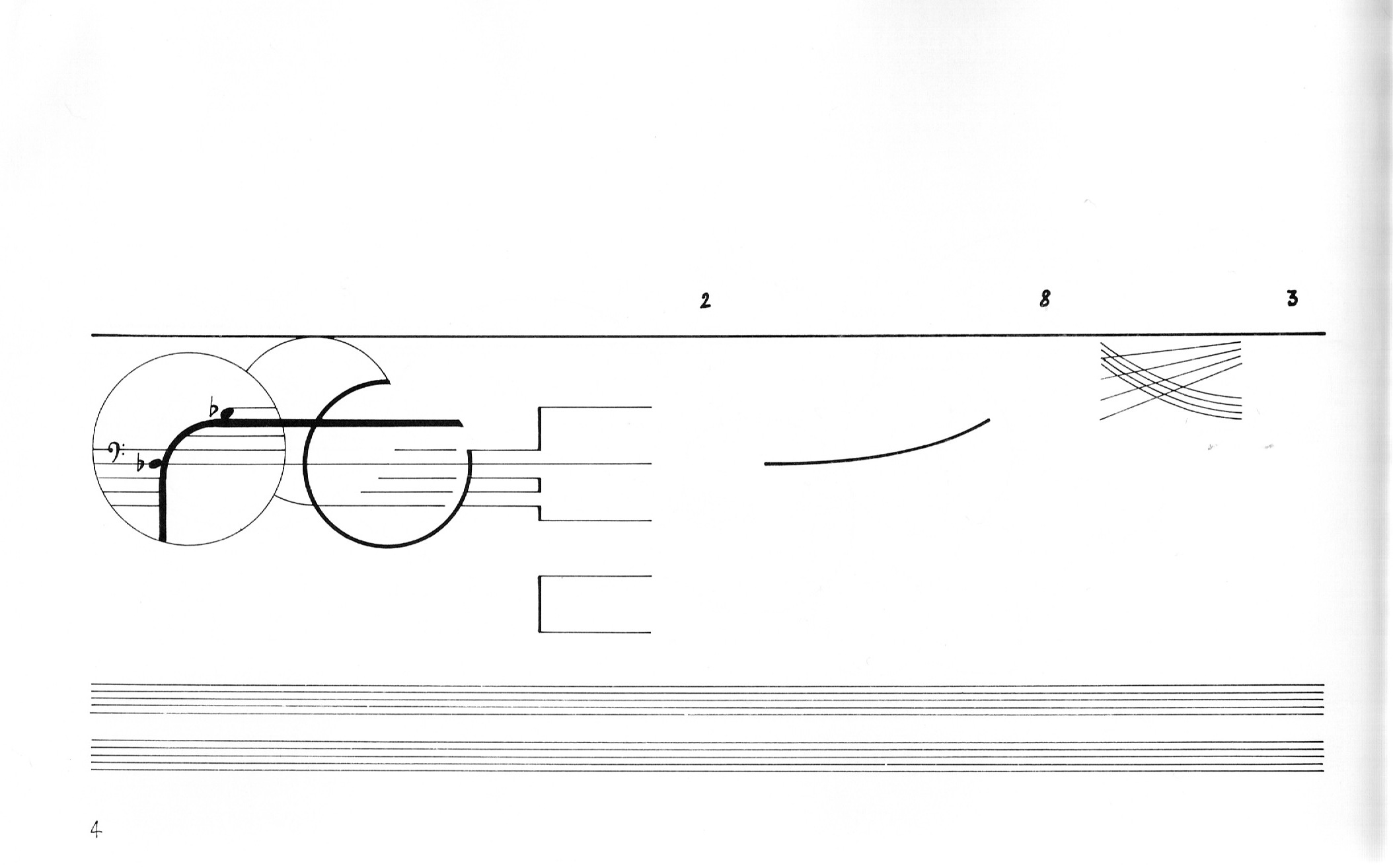}
    \includegraphics[width=\textwidth, height=\textheight, keepaspectratio]{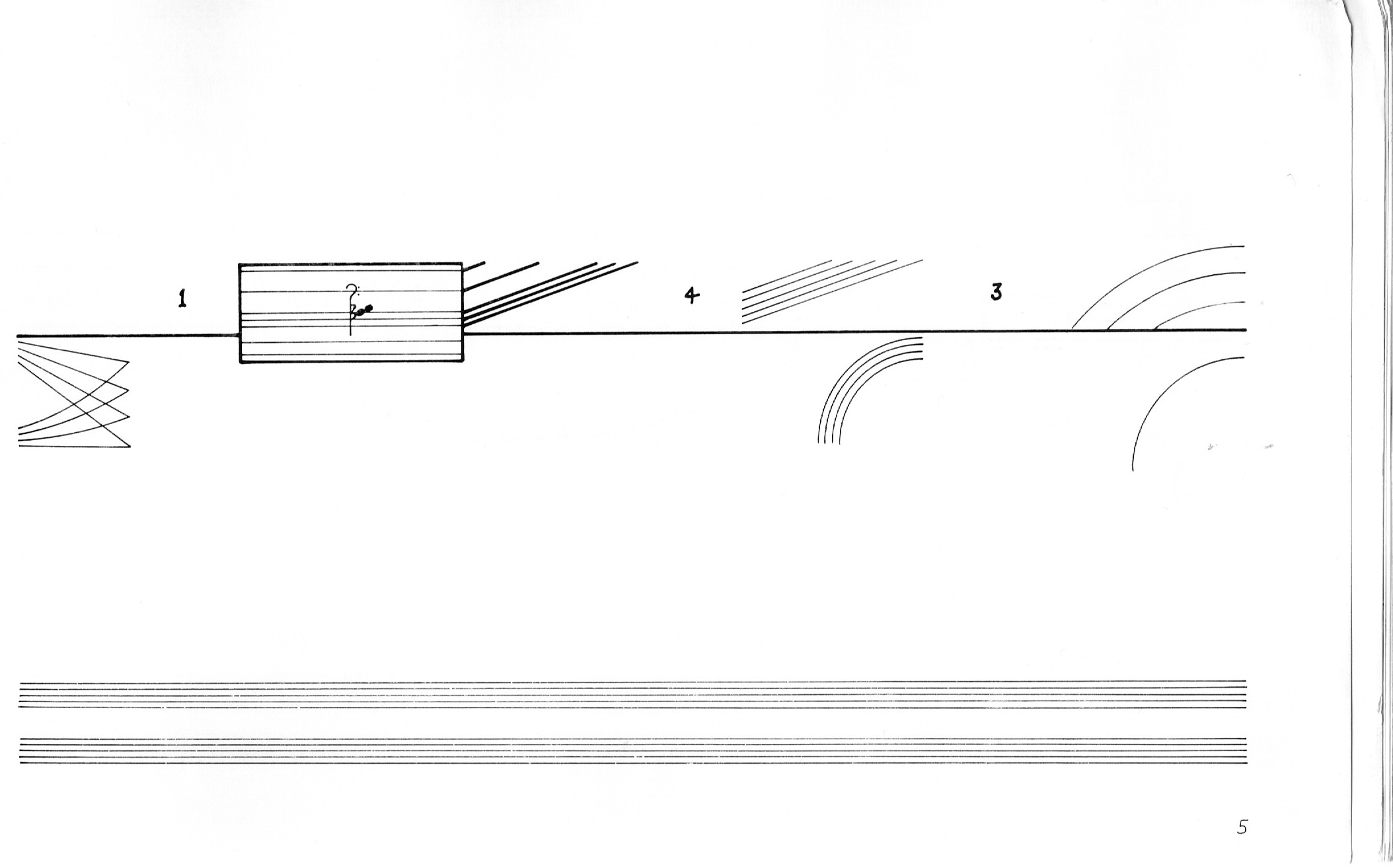}
\end{figure}

\clearpage 
\begin{figure}[!ht] 
    \centering
    \includegraphics[width=\textwidth, height=\textheight, keepaspectratio]{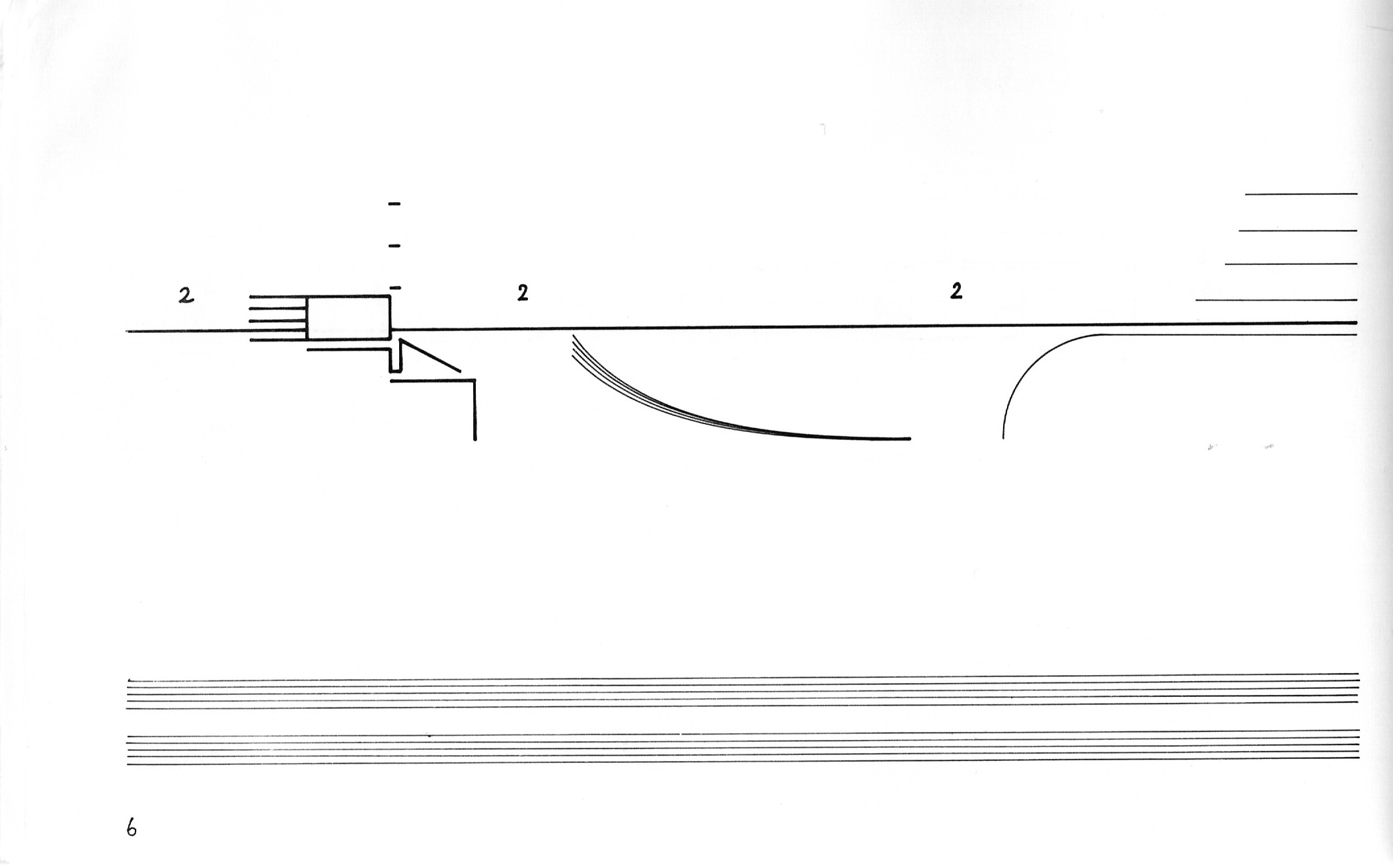}
    \includegraphics[width=\textwidth, height=\textheight, keepaspectratio]{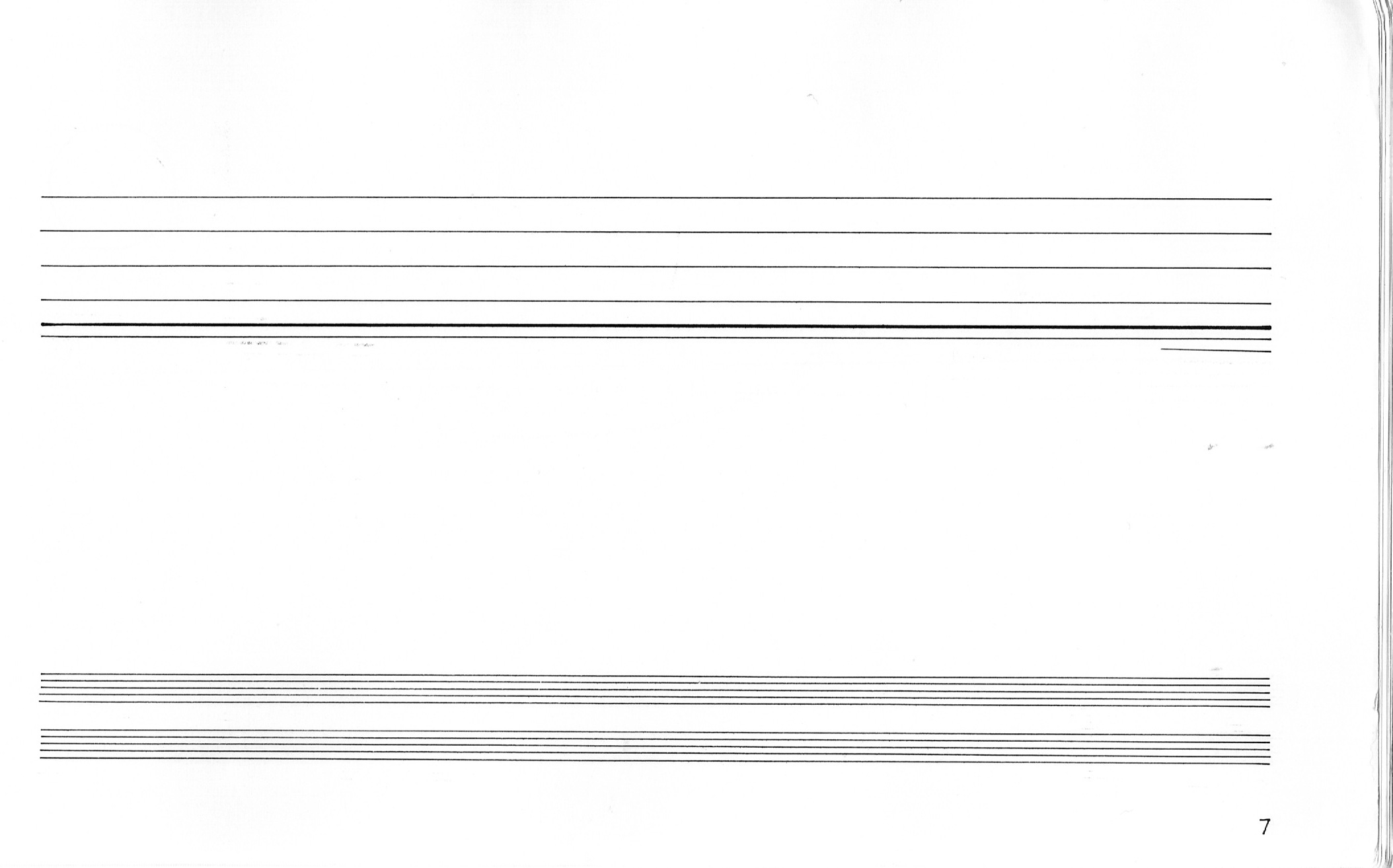}
\end{figure}

\section{Text Prompts pages 1-7}
\label{prompts}

\begin{itemize}
    \item \textbf{Page 1}
    \begin{itemize}
        \item A steady, mid-range tone follows the diagonal line, with a calm, deliberate progression.
        \item Short, rhythmic patterns echo the small horizontal marks, creating subtle variations in texture.
        \item Melodic fragments intertwine with the curves and circles, growing more intricate and layered.
        \item A gentle pulse emerges within the larger circles, gradually softening as the forms expand outward.
    \end{itemize}

    \item \textbf{Page 2}
    \begin{itemize}
        \item A single, continuous tone unfolds steadily, maintaining a calm and balanced presence.
        \item A single, continuous tone unfolds steadily, maintaining a calm and balanced presence.
        \item A single, continuous tone unfolds steadily, maintaining a calm and balanced presence.
        \item The circular shape is met with a warm, resonant tone, expanding and fading into stillness.
    \end{itemize}

    \item \textbf{Page 3}
    \begin{itemize}
        \item A soft, rounded melody mirrors the gentle curve of the left-most circles, expanding outward.
        \item Layers of sound build gradually as more circles overlap, creating a complex and shifting texture.
        \item Sharp, percussive tones cut through the center, reflecting the intersecting lines within the circles.
        \item The sound stretches and smooths out toward the right, softening as the circles become more elongated.
    \end{itemize}

    \item \textbf{Page 4}
    \begin{itemize}
        \item Deep, resonant bass notes emerge from the clef symbol, anchoring the thick, curved lines with weight.
        \item Brief staccato tones interplay between the layered lines, creating sharp rhythmic contrasts.
        \item A slow, sweeping melody rises with the curved line, growing gradually in intensity and pitch.
        \item Interlocking patterns in the crossing lines produce a dense, layered texture, filled with rapid melodic exchanges.
    \end{itemize}

    \item \textbf{Page 5}
    \begin{itemize}
        \item Quick, sharp bursts of sound reflect the interlocking lines on the left, creating a fast-moving rhythm.
        \item Deep, sustained bass notes emerge from the box, anchoring the lines that shoot outward with resonance.
        \item Light, fleeting tones glide along the thin parallel lines, suggesting movement and lightness.
        \item A warm, gradual build in harmony echoes the arches, with each curve adding depth and fullness to the sound.
    \end{itemize}

    \item \textbf{Page 6}
    \begin{itemize}
        \item Short, percussive notes reflect the small, sharp lines on the left, creating a rhythmic start.
        \item A deep, steady tone resonates from the block, grounding the fragmented shapes around it.
        \item Smooth, sliding melodies follow the gentle curve, adding a sense of fluidity and motion.
        \item The upper lines on the right signal a gradual rise in pitch, building toward a light, airy conclusion.
    \end{itemize}

    \item \textbf{Page 7}
    \begin{itemize}
        \item A steady, continuous hum resonates across the horizontal lines, maintaining a calm, balanced texture.
        \item Subtle variations in pitch shift gently between the layers, creating a sense of movement within the stillness.
        \item Tones grow slightly darker and more resonant as the lines thicken, deepening the harmonic texture.
        \item The sound fades softly, becoming lighter and more distant as the thinner lines taper off toward the right.
    \end{itemize}

\end{itemize}

\end{document}